\documentclass{PoS}

\def\simge{\mathrel{%
       \rlap{\raise 0.511ex \hbox{$>$}}{\lower 0.511ex \hbox{$\sim$}}}}
\def\simle{\mathrel{
       \rlap{\raise 0.511ex \hbox{$<$}}{\lower 0.511ex \hbox{$\sim$}}}}

\title{Study of the critical point in lattice QCD at high temperature and density}

\ShortTitle{Study of the critical point in lattice QCD at high temperature and density}

\author{\speaker{Shinji Ejiri}%
\thanks{This work has been authored under contract DE-AC02-98CH1-886
with the U.S. Department of Energy.}
\\
Physics Department, Brookhaven National Laboratory,
Upton, New York 11973, USA\\
        E-mail: \email{ejiri@quark.phy.bnl.gov}}

\abstract{
We propose a method to probe the nature of phase transitions in lattice 
QCD at finite temperature and density, which is based on the investigation 
of an effective potential as a function of the average plaquette. 
We analyze data obtained in a simulation of two-flavor QCD using 
p4-improved staggered quarks with bare quark mass $m/T = 0.4$, 
and find that a first order phase transition line appears in 
the high density regime for $\mu_q/T \simge 2.5$. 
The effective potential as a function of the quark number density 
is also studied. 
We calculate the chemical potential as a function of the density 
from the canonical partition function and discuss the existence of 
the first order phase transition line. 
}

\FullConference{The XXV International Symposium on Lattice Field Theory\\
		July 30 - August 4 2007\\
		Regensburg, Germany}

\begin{document}

\section{Effective potential as a function of the average plaquette}

The study of the QCD phase diagram at non-zero temperature $(T)$ and 
chemical potential $(\mu_q)$ is one of the most important topics 
among studies of lattice QCD.
In particular, the study of the endpoint of the first order phase 
transition line in the $(T, \mu_q)$ plane is particularly interesting 
both from the experimental and theoretical point of view. 
The existence of such a critical point is suggested by phenomenological 
studies. The appearance of the critical endpoint in the $(T, \mu_q)$ plane 
is closely related to hadronic fluctuations in heavy ion collisions 
and may be experimentally examined by an event-by-event analysis 
of heavy ion collisions.
Many trials have been made to prove the existence of the critical 
endpoint by first principle calculation in lattice QCD, 
however no definite conclusion on this issue is obtained so far.
The purpose of this study is to clarify the existence of the endpoint 
of the first order phase transition line in the $(T, \mu_q)$ plane. 
We propose a new method to investigate the nature of transition.

We evaluate an effective potential as a function of 
the average plaquette $(P)$
\footnote{For later discussions, we define the average plaquette $P$ as 
$ P \equiv -S_g/(6 \beta N_{\rm site})$.
This is the average of the plaquette over all elementary squares 
for the standard gauge action.}, and identify the type of 
transition from the shape of the potential. 
The partition function can be written as 
\footnote{We restrict ourselves to discuss only the case when the quark 
matrix does not depend on $\beta$ explicitly, e.g. the standard 
Wilson and staggered quark actions, the p4-improved staggered 
quark action etc., for simplicity.}
\begin{eqnarray}
{\cal Z}_{\rm GC}(\beta, \mu_q) = \int 
{\cal D}U \left( \det M(\mu_q) \right)^{N_{\rm f}} e^{-S_g(P,\beta)}
= \int R(P,\mu_q) w(P) e^{-S_g(P,\beta)} \ dP,
\label{eq:rewmuP}
\end{eqnarray}
where $S_g(P,\beta)$ is the gauge action, 
$w(P)$ is the state density at $\mu_q=0$ for each $P$, 
\begin{eqnarray}
w(P')e^{-S_g(P',\beta)} \equiv 
w(P',\beta) \equiv \int {\cal D} U \ \delta(P'-P) \ (\det M)^{N_{\rm f}} 
e^{6\beta N_{\rm site} P}, 
\label{eq:pdist}
\end{eqnarray}
and $R(P,\mu_q)$ is the modification (reweighting) factor for finite 
$\mu_q$, which is defined by 
\begin{eqnarray}
R(P',\mu_q) \equiv 
\frac{\int {\cal D} U \ \delta(P'-P) (\det M(\mu_q))^{N_{\rm f}}}{
\int {\cal D} U \ \delta(P'-P) (\det M(0))^{N_{\rm f}}} 
= \frac{ \left\langle \delta(P'-P) 
\left( \det M(\mu_q) / \det M(0) \right)^{N_{\rm f}}
\right\rangle_{(\beta, \mu_q=0)} }{
\left\langle \delta(P'-P) \right\rangle_{(\beta, \mu_q=0)}},
\label{eq:rmudef}
\end{eqnarray}
where $\left\langle \cdots \right\rangle_{(\beta, \mu_q=0)}$ means 
the expectation value at $\mu_q=0$, 
$\det M$ is the quark determinant, 
$N_{\rm f}$ is the number of flavors
\footnote{$N_{\rm f}$ must be replaced in these equations to $N_{\rm f}/4$ 
when we use a staggered type quark action.} , and
$N_{\rm site} = N_s^3 \times N_t$ is the number of sites.
We then define the effective potential as 
$V(P, \beta, \mu_q) =- \ln (R w e^{-S_g})$. 
If there is a first order phase transition point, where two different 
states coexist, the potential must have two 
minima at two different values of $P$. 
In this paper, we discuss whether the potential at $\mu_q=0$, i.e. 
$\ln(w e^{-Sg})$, which is quadratic function when the transition is 
a crossover, can change to a double-well potential by the reweighting 
factor for finite $\mu_q$, as illustrated in Fig.~\ref{fig1} (left).

\section{Taylor expansion in terms of $\mu_q/T$ and Gaussian distribution}

\begin{figure}[t]
\begin{center}
\includegraphics[width=2.2in]{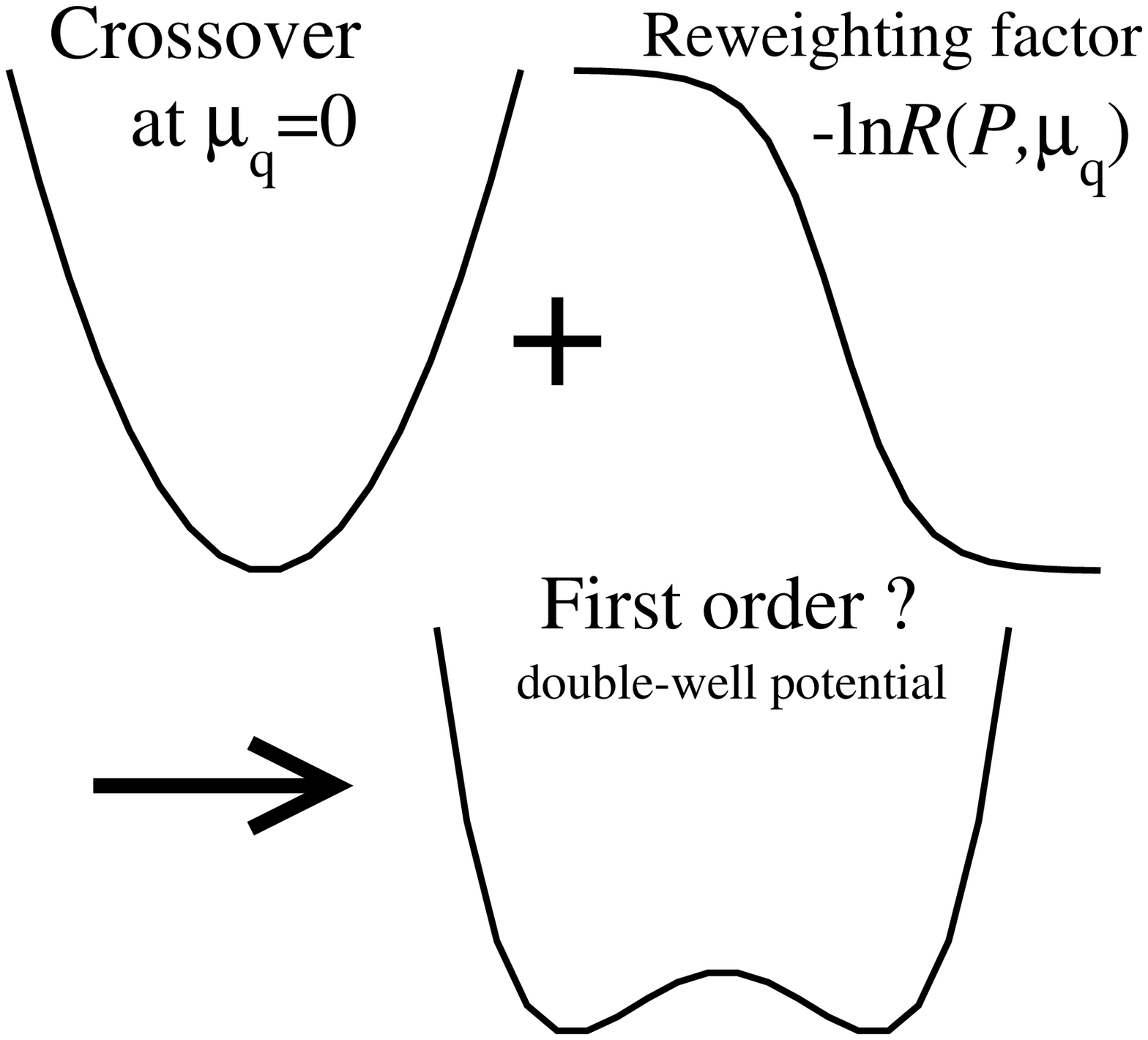}
\hskip 1.5cm
\includegraphics[width=2.7in]{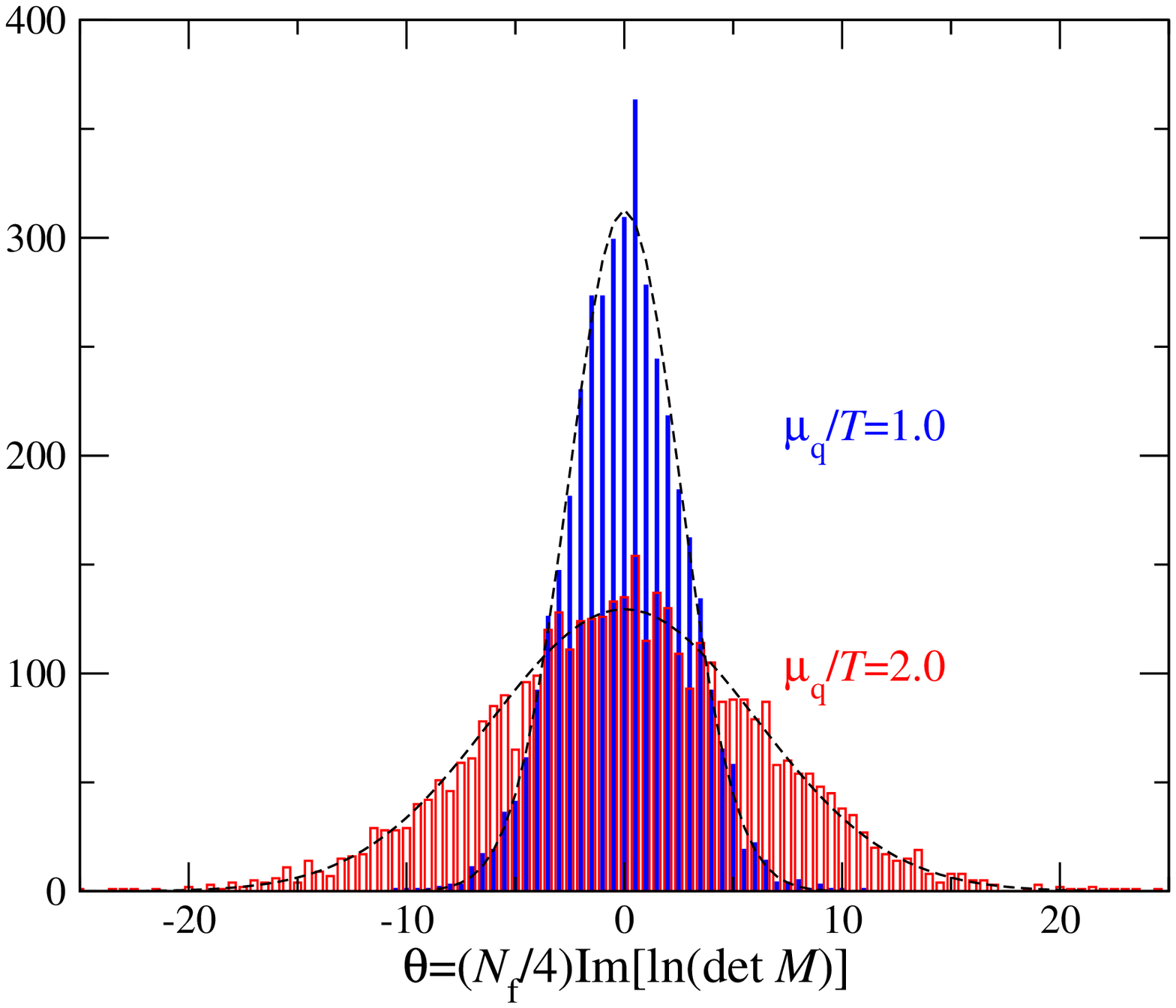}
\vskip -0.2cm
\caption{(left) Schematic figures of the effective potential and 
the reweighting factor. \ 
(right) The histogram of the complex phase for $\mu_q/T=1.0$ 
and $2.0$ at $\beta=3.65$.
}
\label{fig1}
\end{center}
\vskip -0.3cm
\end{figure} 

However, the calculation of the quark determinant is quite expensive 
and is actually difficult except on small lattices. Moreover, 
the calculation of $R(P,\mu_q)$ becomes increasingly more difficult 
for large $\mu_q$ due to the sign problem, i.e. the statistical error 
becomes exponentially larger as $\mu_q$ increases.
We avoid these problems by the following two ideas. 
One is that we perform a Taylor expansion of $\ln \det M(\mu_q)$ 
in terms of $\mu_q$ at $\mu_q=0$ and calculate the expansion 
coefficients \cite{BS02}, 
\begin{eqnarray}
\ln \left[ \frac{\det M(\mu_q)}{\det M(0)} \right] 
= \sum_{n=1}^{\infty} \frac{1}{n!} 
\left[ \frac{\partial^n (\ln \det M)}{\partial (\mu_q/T)^n} \right] 
\left( \frac{\mu_q}{T} \right)^n 
\equiv N_s^3 N_t \sum_{n=1}^{\infty} D_n \left( \frac{\mu_q}{T} \right)^n . 
\label{eq:detTay}
\end{eqnarray}
The Taylor expansion coefficients are rather easy to calculate 
by using the stochastic noise method. 
Although we must cut off this expansion at an appropriate order 
in $\mu_q$, we can estimate the application range where 
the approximation is valid for each analysis \cite{BS03,BS05}. 
While the application range of the Taylor expansion of 
$\ln {\cal Z}_{\rm GC}$ should be limited by the critical point 
because $\ln {\cal Z}_{\rm GC}$ is singular at the critical point, 
there is no such limit for the application range in the expansion 
of $\ln R(P,\mu_q)$ 
because the weight factor should always be well-defined. 

The sign problem is avoided by the following idea.
We consider a probability distribution function as a function of 
the complex phase of the quark determinant $\theta$, 
$|F| \equiv |\det M(\mu_q)/\det M(0)|^{N_{\rm f}}$ and $P$, 
i.e. $\bar{w} (P, |F|, \theta)$.
If we assume the distribution function in $\theta$ is well-approximated 
by a Gaussian function, the sign problem in the calculation of 
$\ln R(P,\mu_q)$ is completely solved \cite{Eji07}. 

We define the complex phase by a Taylor expansion, 
$\theta = N_{\rm f}N_s^3N_t \sum_{n=0}^{\infty} 
{\rm Im} D_{2n+1} (\mu_q /T)^{2n+1}$. 
Since the partition function is real even at non-zero density, 
the distribution function has the symmetry under 
the change from $\theta$ to $-\theta$.
Therefore, the distribution function is written by 
$\bar{w}(\theta) \sim \exp[-(a_2 \theta^2 +a_4 \theta^4 
+a_6 \theta^6 + \cdots)].$
Moreover, because $D_n$ is a trace of a matrix which has space 
index \cite{BS05}, 
e.g. $D_1 \propto {\rm tr}[ M^{-1} dM/d(\mu_q/T)]$, 
the central limit theorem suggests that the distribution 
function is well-approximated by a Gaussian function, 
when the system size is sufficiently large in comparison to 
the correlation length between diagonal elements of the matrix. 
We plotted the distribution of the complex phase in Fig.~\ref{fig1} 
(right) and fitted by a Gaussian function (dashed line). 
It is found that this approximation is quite well, 
hence we consider the leading term of the expansion only, 
$\bar{w}(P, |F|, \theta) \propto \sqrt{a_2 (P, |F|)/\pi} \ 
\exp [-a_2 (P, |F|) \theta^{2} ]. $
The coefficient $a_2 (P, |F|)$ is given by 
$1/(2a_2)=\left\langle \theta^2 \right\rangle$
for each $P$ and $|F|$.
The numerator of Eq.~(\ref{eq:rmudef}) is then evaluated by 
\begin{eqnarray}
\left\langle F(\mu_q) \delta(P'-P) \right\rangle_{(\beta, \mu_q=0)}
\approx \left\langle e^{-1/(4a_2(P, |F|))} 
| F(\mu_q) | \delta(P'-P) \right\rangle_{(\beta, \mu_q=0)}.
\label{eq:den12}
\end{eqnarray}
Because $1/a_2 \sim O(N_{\rm site})$, the phase factor in $R(P,\mu_a)$ 
decreases exponentially as a function of the volume. 
However, the operator in Eq.~(\ref{eq:den12}) is always real and positive 
for each configuration in this framework, hence the expectation value 
of $R(P, \mu_q)$ is always larger than its statistical error, namely 
the contribution $\ln R(P,\mu_q)$ to the effective potential 
$V(P, \beta, \mu_q)$ is always well-defined.
Therefore, the sign problem is completely avoided if we can assume 
the Gaussian distribution of $\theta$.

\section{Numerical results of the effective potential}

\begin{figure}[t]
\begin{center}
\includegraphics[width=2.7in]{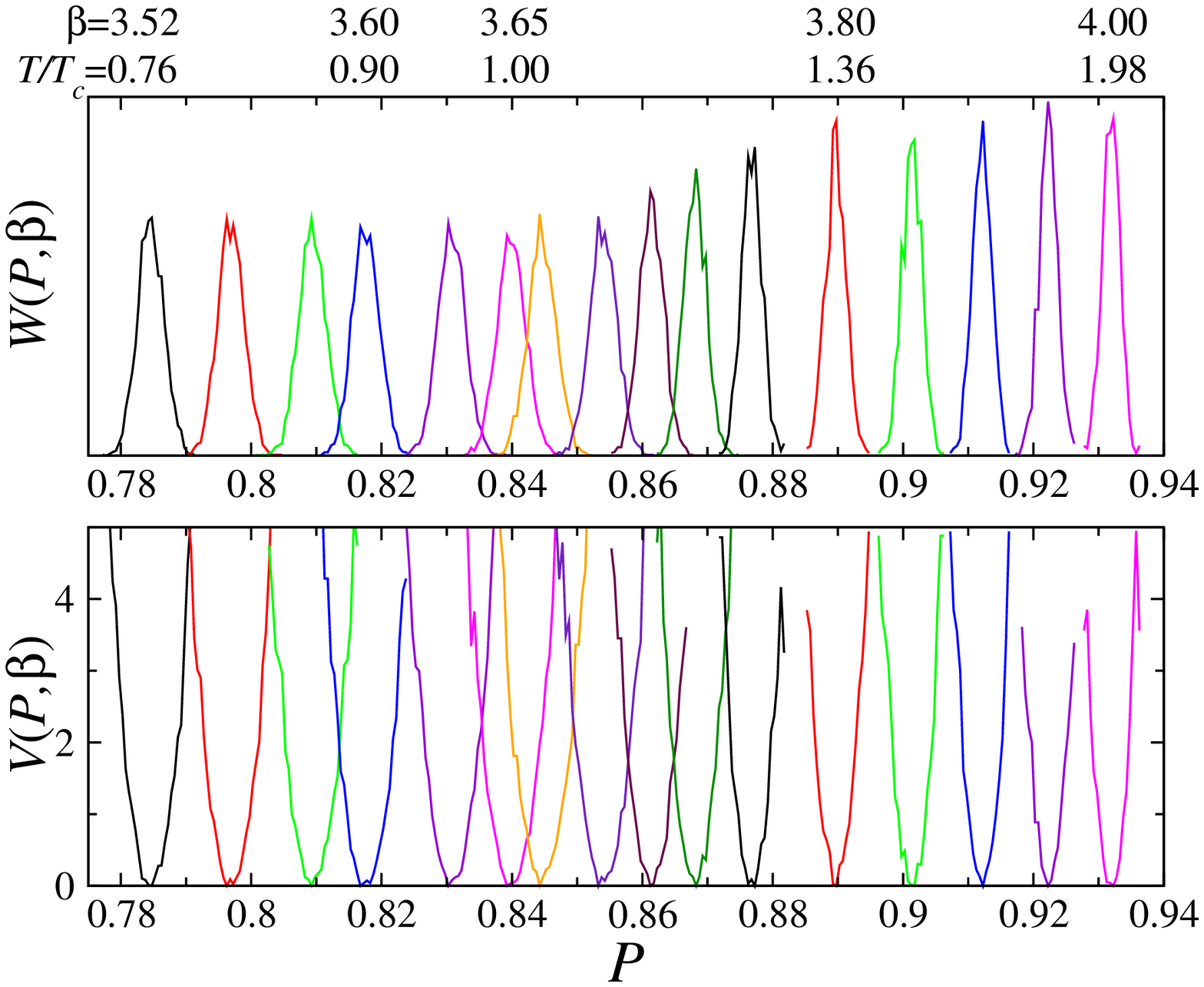}
\hskip 0.5cm
\includegraphics[width=2.7in]{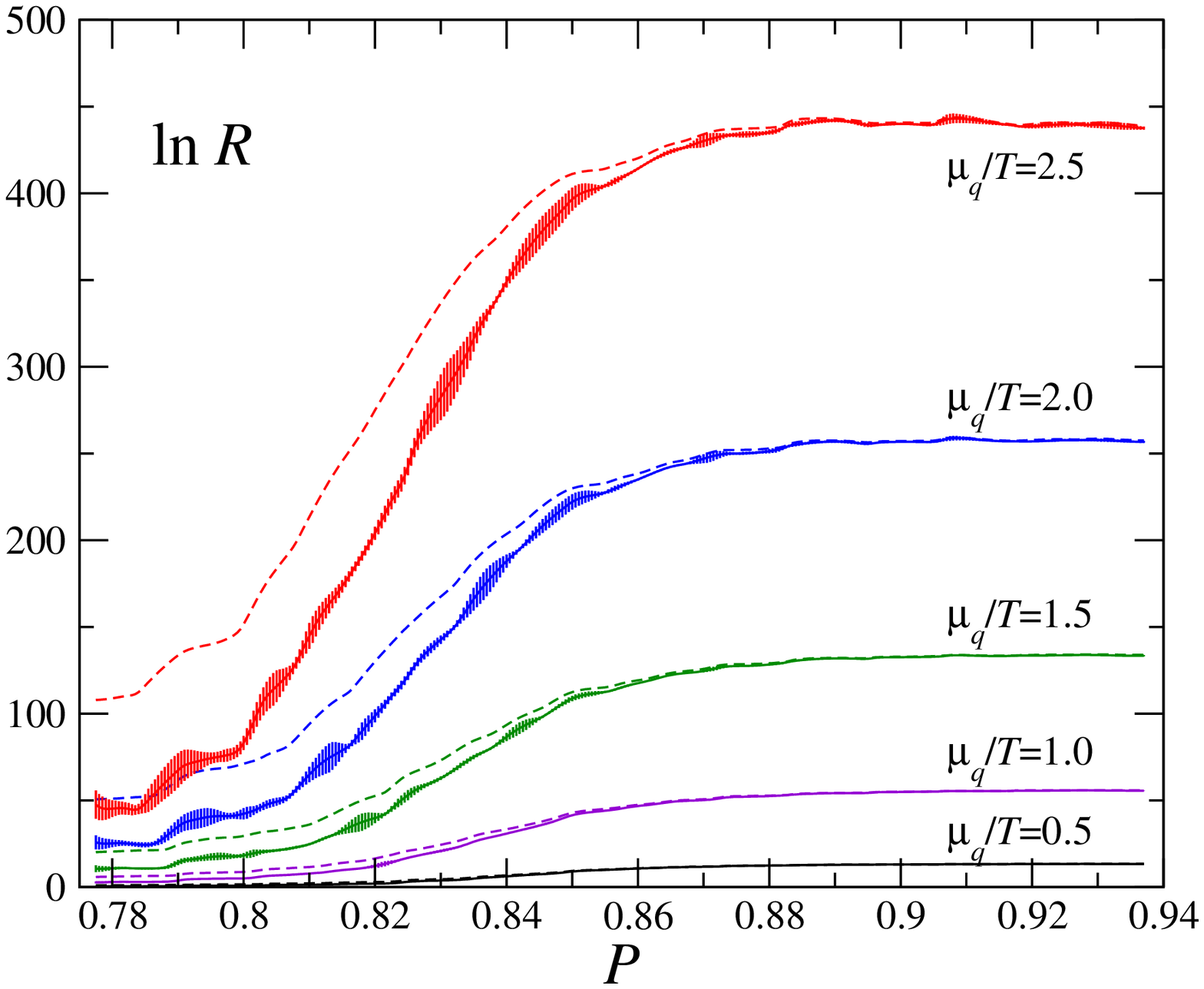}
\vskip -0.2cm
\caption{(left) The plaquette histogram and the effective potential 
at $\mu_q=0$. \ 
(right) The reweighting factor $\ln R (P, \mu_q)$ for $\mu_q/T=0.5 - 2.5$.
}
\label{fig2}
\end{center}
\vskip -0.3cm
\end{figure} 

We calculate $w(P,\beta)$ and $R(P,\mu_q)$ using data obtained 
by simulations in \cite{BS05}.
The Taylor expansion coefficients are computed up to $O(\mu_q^6)$. 
These are measured at sixteen simulation points from $\beta=3.52$ 
to $4.00$ for the bare quark mass $ma=0.1$. 
The corresponding temperature normalized by the pseudo-critical 
temperature is in the range of $T/T_c= 0.76$ to $1.98$. 
The ratio of pseudo-scalar and vector meson masses is 
$m_{\rm PS}/m_{\rm V} \approx 0.7$ at $\beta=3.65$.
The lattice size $N_{\rm site}$ is $16^3 \times 4$. 
The number of configurations is 1000 -- 4000 for each $\beta$. 

The probability distribution function $w(P,\beta)$, 
i.e. the histogram of $P$ for each $\beta$, and the effective potential 
$V(P,\beta)$ at $\mu_q=0$ are given in Fig.~\ref{fig2}(left). 
To obtain $w(P,\beta)$, 
we grouped the configurations by the value of $P$ into blocks 
and counted the number of configurations in these blocks, 
and the potential $V(P,\beta)$ is normalized by the minimum value 
for each temperature.

The results for $\ln R(P, \mu_q)$ are shown by solid lines in 
Fig.~\ref{fig2} (right) for $\mu_q/T=0.5, 1.0, 1.5, 2.0$ and $2.5$. 
We find a rapid change in $\ln R$ around $P \sim 0.83$, and 
the variation becomes larger as $\mu_q/T$ increases. 
The dashed lines in Fig.~\ref{fig2} (right) are the results that 
we obtained when the effect of the complex phase,
i.e. $\exp[-1/(4a_2)]$, is omitted. 
These dashed lines correspond to the reweighting 
factor with non-zero isospin chemical potential $\mu_I$ and 
zero quark chemical potential $\mu_q$ \cite{KS06}.
The variation of $\ln R$ in terms of $P$ becomes milder 
when the effect of the complex phase is omitted. 
This explains the difference between the phase diagrams of QCD 
with non-zero quark chemical potential and non-zero isospin 
chemical potential \cite{Eji07}.

\begin{figure}[t]
\begin{center}
\includegraphics[width=2.7in]{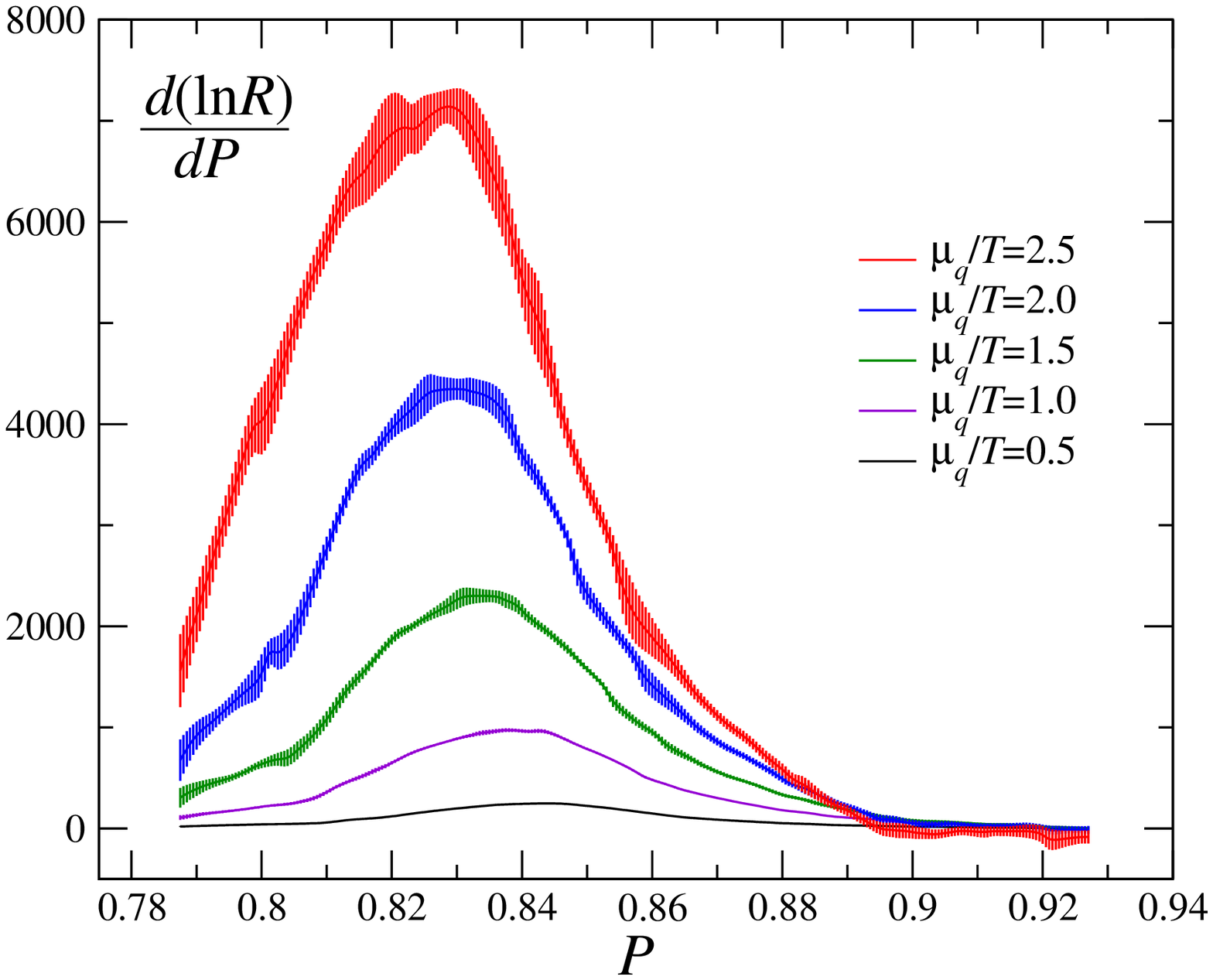}
\hskip 0.5cm
\includegraphics[width=2.7in]{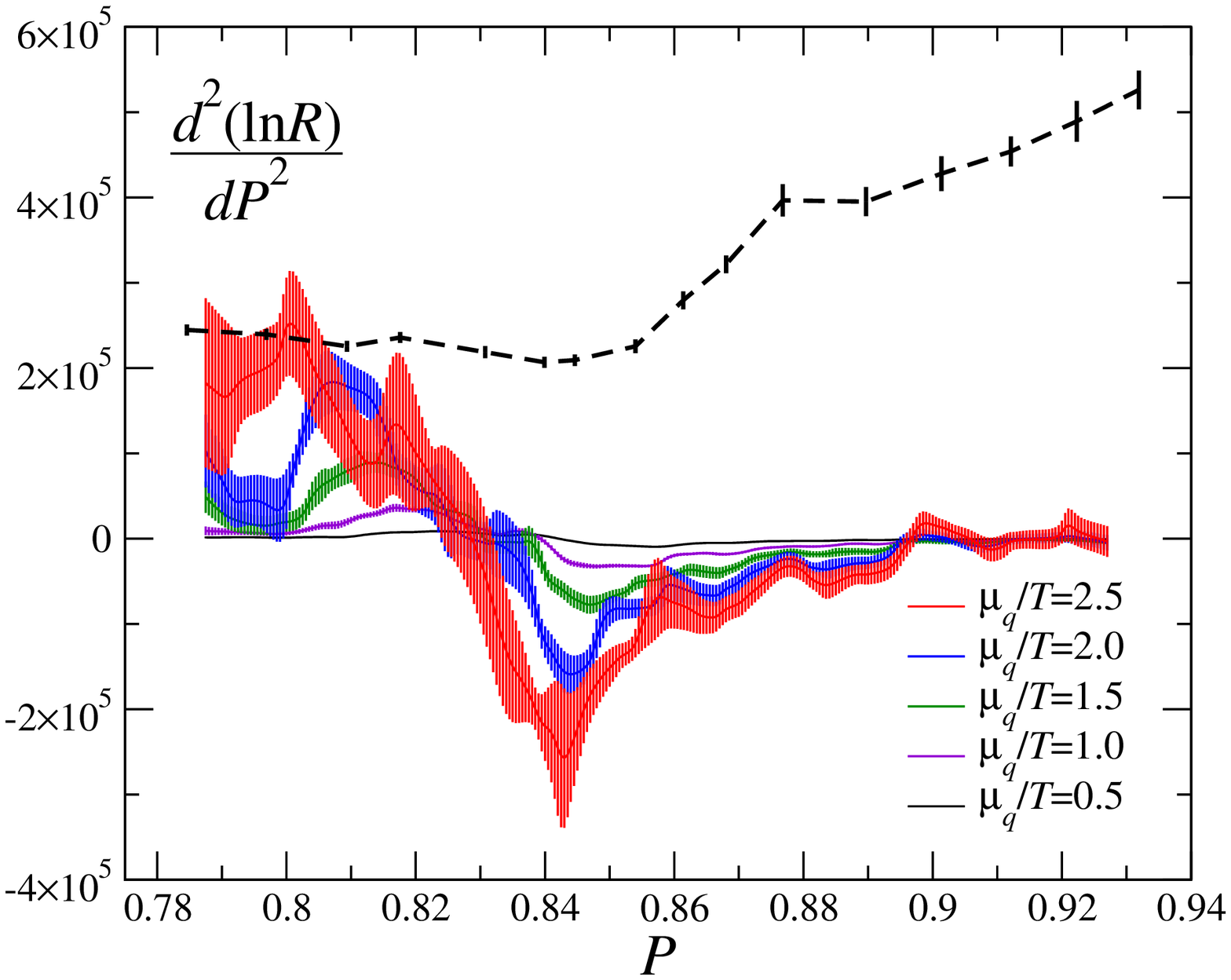}
\vskip -0.2cm
\caption{The slope (left) and curvature (right) of $\ln R(P,\mu_q)$.
The dashed line is the curvature of $- \ln w$.
}
\label{fig3}
\end{center}
\vskip -0.3cm
\end{figure}

We discuss the shape of the effective potential at non-zero $\mu_q$.
The effective potential is obtained from 
$V(P, \beta, \mu_q)=-\ln w(P,\beta) -\ln R(P,\mu_q)$ 
substituting the data in Fig.~\ref{fig2}. 
From Eq.~(\ref{eq:pdist}), the slope of $V(P, \beta, \mu_q)$ can 
be controlled by $\beta$, 
\begin{eqnarray}
V(P,\beta,\mu_q) = V(P,\beta_0,\mu_q) -6 (\beta - \beta_0) N_{\rm site} P.
\label{eq:vrewbeta}
\end{eqnarray}
under a change of $\beta_0 \to \beta$, however the curvature of 
the potential does not change by $\beta$. 
We expect that the curvature vanishes at the endpoint of the first 
order phase transition line by canceling $d^2(\ln w)/dP^2$ and 
$d^2(\ln R)/dP^2$.
In order to analyze the sign of $d^2V/dP^2(P, \mu_q)$, 
we fitted the data of $\ln R$ by a quadratic function of $P$
and calculate the first and second derivatives of $\ln R(P, \mu_q)$ 
at each $P$. 
The results of the slope and curvature are 
shown in Fig.~\ref{fig3} for each $\mu_q/T$. 
In the region around $P \sim 0.83$, $d (\ln R)/dP$ 
becomes larger as $\mu_q/T$ increases and $\ln R(P, \mu_q)$ 
changes sharply in this region.
The magnitude of the curvature of $\ln R$ also becomes larger 
as $\mu_q/T$ increases.

To evaluate $d^2 (\ln w)/dP^2(P)$, we assume $w(P,\beta)$ in 
Fig.~\ref{fig2} (left) is a Gaussian function. In this case, 
the curvature at $\langle P \rangle$ is given by 
$-d^2 (\ln w)/dP^2 = 6N_{\rm site}/\chi_P, $
where 
$\chi_P \equiv 6N_{\rm site} \langle (P - \langle P \rangle)^2 \rangle$ 
is the plaquette susceptibility. 
The dashed line in Fig.~\ref{fig3} (right) is the result of 
$-d^2 (\ln w)/dP^2(P)$. 

It is found from Fig.~\ref{fig3} (right) that the maximum value of 
$d^2 (\ln R)/dP^2 (P, \mu_q)$ at $P=0.80$ becomes larger 
than $-d^2 (\ln w)/dP^2$ for $\mu_q/T \simge 2.5$. 
This means that the curvature of the effective potential 
vanishes at $\mu_q/T \sim 2.5$ 
and becomes negative for large $\mu_q/T$, 
namely the shape of the effective potential which is of quadratic 
type at $\mu_q=0$ changes to a double-well type at large $\mu_q/T$. 
For the quantitative estimation of the endpoint of the first order 
phase transition, further investigation must be needed. 
However, this argument strongly suggests the existence of 
the first order phase transition line in the $(T, \mu_q)$ plane. 
Further details of this analysis are given in \cite{Eji07}.

\section{Canonical partition function}

\begin{figure}[t]
\begin{center}
\includegraphics[width=2.7in]{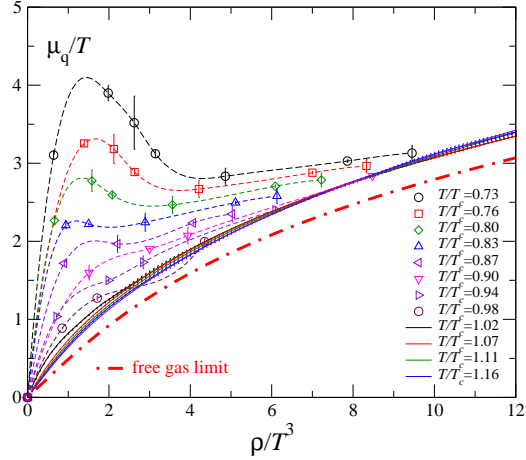}
\vskip -0.2cm
\caption{
The chemical potential as a function of the quark number density.
}
\label{fig4}
\end{center}
\vskip -0.3cm
\end{figure} 

Next, we want to apply the effective potential argument to 
the weight factor as a function of the quark number density $\rho$. 
The physical meaning of this potential is clearer than that of $P$ 
because the weight factor for each quark number $N$ corresponds 
to the canonical partition function ${\cal Z}_{\rm C}$, 
\begin{eqnarray}
{\cal Z}_{\rm GC}(T,\mu_q) 
= \sum_{N} \ e^{N \mu_q/T} {\cal Z}_{\rm C}(T,N), \hspace{3mm}
N=\bar{\rho} N_s^3, \hspace{3mm}
\rho/T^3=\bar{\rho} N_t^3.
\label{eq:partition} 
\end{eqnarray}

The canonical partition function can be given by an inverse Laplace 
transformation \cite{Mill87,Hase92,Alex05},
\begin{eqnarray}
{\cal Z}_{\rm C}(T,N) = \frac{3}{2 \pi} \int_{-\pi/3}^{\pi/3} d(\mu_I/T)
e^{-N (\mu_0/T+i\mu_I/T)} {\cal Z}_{\rm GC}(T, \mu_0+i\mu_I),
\label{eq:canonicalP} 
\end{eqnarray}
where $\mu_0$ is an appropriate real constant. 
Note that 
${\cal Z}_{\rm GC}(T, \mu_q +2\pi iT/3) = {\cal Z}_{\rm GC}(T, \mu_q)$.
Recently, this canonical partition function is calculated for 
$N_{\rm f}=4$ using the Glasgow method \cite{Krat05}. 
However, with present day computer resources, the calculation by 
the Glasgow method is difficult except on small lattices. 
We consider approximations which is valid for large volume and low density 
in this approach, as discussed in the first part of this paper.

We calculate the grand partition function by the Taylor 
expansion, Eq.~(\ref{eq:detTay}), 
\begin{eqnarray}
\frac{{\cal Z}_{\rm GC}(T, \mu_q)}{{\cal Z}_{\rm GC}(T,0)}
= \frac{1}{{\cal Z}_{\rm GC}} \int {\cal D}U 
\left( \frac{\det M(\mu_q)}{\det M(0)} \right)^{N_{\rm f}}
(\det M(0))^{N_{\rm f}} e^{-S_g} 
\equiv 
\left\langle e^{\left[ N_{\rm f} N_t V \sum_{n=1}^{\infty} D_n 
\left( \frac{\mu_q}{T} \right)^n \right]} \right\rangle_{(T, \mu_q=0),}
\label{eq:normZGC} 
\end{eqnarray}
where $V \equiv N_s^3$. We moreover use a saddle point approximation, 
which is valid for a large system. 
We find a saddle point $z_0$ in the complex $\mu_q/T$ plane for 
each configuration, which satisfies
$ \left[ N_{\rm f} N_t \sum_{n=1}^{\infty} n D_n z^{n-1} 
- \bar{\rho} \right]_{z=z_0} =0$. 
The canonical partition function is given by 
\begin{eqnarray}
{\cal Z}_{\rm C}(T, \bar{\rho} V) 
\approx \frac{3}{\sqrt{2 \pi}} {\cal Z}_{\rm GC} (T,0)
\left\langle \exp \left[ V \left( N_{\rm f} N_t 
\sum_{n=1}^{\infty} D_n z_0^n - \bar{\rho} z_0 \right) \right] 
e^{-i \alpha/2} \sqrt{ \frac{1}{V |D''(z_0)|}}
\right\rangle_{(T, \mu=0)}
\label{eq:zcspa}
\end{eqnarray}
for large $V$. 
Here, $ D''(z) = (d^2/dz^2) 
\left( N_{\rm f} N_t \sum_{n=1}^{\infty} D_n z^n \right) $
and $D''(z_0)=|D''(z_0)| e^{i \alpha}$.

The chemical potential, i.e. the slope of 
the effective potential, is also evaluated by 
\begin{eqnarray}
\frac{\mu_q}{T} 
\ = \ \frac{-1}{V} \frac{\partial \ln {\cal Z}_C (T, \bar{\rho} V)}
{\partial \bar{\rho}}
\approx \frac{
\left\langle z_0 \exp \left[ V \left( N_{\rm f} N_t 
\sum_{n=1}^{\infty} D_n z_0^n - \bar{\rho} z_0 \right) \right] 
e^{-\frac{i \alpha}{2}} \sqrt{ \frac{1}{V |D''(z_0)|}}
\right\rangle_{(T, \mu_q=0)}}{
\left\langle \exp \left[ V \left( N_{\rm f} N_t 
\sum_{n=1}^{\infty} D_n z_0^n - \bar{\rho} z_0 \right) \right] 
e^{-\frac{i \alpha}{2}} \sqrt{ \frac{1}{V |D''(z_0)|}}
\right\rangle_{(T, \mu_q=0)}}. 
\label{eq:chem}
\end{eqnarray}
This equation is similar to the formula of the reweighting method 
for finite $\mu_q$. 
The operator in the denominator corresponds to a reweighting factor, 
and the chemical potential is an expectation value of the saddle 
point calculated with this modification factor.

We analyze the data used in the previous section. 
The Taylor expansion coefficients up to $O(\mu_q^6)$ are used. 
The volume $V=16^3$ would be sufficiently large, and we assume a Gaussian 
distribution function for the complex phase of the reweighting factor, again.
We find a saddle point $z_0$ numerically for each configuration, 
assuming $z_0$ exists near the real axis in the low density region of 
the complex $\mu_q/T$ plane. 
We use multi-$\beta$ reweighting method \cite{Swen88} combining all data 
obtained at 16 points of $\beta$. 
Configurations are generated with the provability of the Boltzmann weight 
in Monte-Carlo simulations, however the important configurations will 
change when the weight is changed by the reweighting method. 
For such a case, the multi-$\beta$ reweighting is effective, since 
the important configurations are automatically selected among all 
configurations generated at multi-$\beta$, and also this method is useful 
for the interpolation between the simulation points. 

We plot the result of $\mu_q/T$ in Fig.~\ref{fig4} as a function of 
$\rho/T^3$ for each temperature. 
The dot-dashed line is the value in the free gas limit.
As seen in Fig.~\ref{fig2} (left), the configurations do not 
distribute uniformly in the range of $P$ which is necessary in 
this analysis, and correct results cannot be obtained if the important 
configurations are missing. 
At low temperature, the important value of $P$ changes very much as $\rho$ 
increases, therefore we plotted only the data when the expectation value 
of $P$ is on the peaks of the histograms of $P$ in Fig.~\ref{fig2} (left). 
The dashed lines are cubic spline interpolations of these data.

It is found from this figure that a qualitative feature of $\mu_q/T$ 
changes around $T/T_c \sim 0.8$, i.e. $\mu_q/T$ increases monotonically 
as $\rho$ increases above 0.8, whereas it shows an s-shape below 0.8. 
This means that there is more than one values of $\rho/T^3$ for 
one value of $\mu_q/T$ below $T/T_c \sim 0.8$.
This is a signature of a first order phase transition. 
The critical value of $\mu_q/T$ is about $2.5$, which is consistent 
with the result in the previous section. 
Although further studies including justifications of these 
approximations used in this analysis are necessary for more 
qualitative investigation, this result also suggests the existence of 
the first order phase transition line in the $(T, \mu_q)$ plane.


\end{document}